\documentclass{article}

\usepackage[english]{babel}

\usepackage[letterpaper,top=2cm,bottom=2cm,left=3cm,right=3cm,marginparwidth=1.75cm]{geometry}

\usepackage{amsmath}
\usepackage{graphicx}
\usepackage[colorlinks=true, allcolors=blue]{hyperref}

\title{AudioWorldSim: Realistic Binaural Audio Datasets For World Models}
\author{
    Luis Vitor Zerkowski \\
    VISGRAF \\
    IMPA \\
    \texttt{luisvz@gmail.com}
    \and
    Luiz Velho \\
    VISGRAF \\
    IMPA \\
    \texttt{lvelho@impa.br}
}

\begin{document}
\maketitle

\begin{abstract}
    This technical report presents AudioWorldSim, an open-source platform designed to generate realistic binaural audio datasets and advance research in audio-based machine learning, particularly world models. Built as a custom extension of Meta's SoundSpaces 2.0 platform, AudioWorldSim leverages their comprehensive acoustics framework, but focuses on the automatic rollout of random agent navigations, as well as implements crucial fixes to how continuous sound is composed. AudioWorldSim is made publicly available to the research community at https://github.com/Luizerko/AudioWorldSim to facilitate reproducibility.
\end{abstract}

\section{Introduction}

    World models have emerged as a central paradigm in artificial intelligence, enabling agents to internalize environment dynamics, predict future states, and plan actions within a learned latent space \cite{ha2018worldmodels}. While early foundational architectures focused predominantly on visual signals \cite{hafner2024masteringdiversedomainsworld,assran2025vjepa2selfsupervisedvideo,bar2025navigationworldmodels}, human perception relies heavily on multimodal integration. Audio provide essential environmental information, including event localization, occlusions, and even material composition. Integrating binaural spatial audio into world models allows artificial agents to achieve a more comprehensive world state representation, moving toward true multisensory imagination \cite{wang2026audiovisualworldmodelslearning}.

    Achieving this requires accurate acoustic information, yet there is a severe scarcity of high-quality, realistic, binaural data tailored for spatial navigation rather than speech. The few available datasets -- particularly those that are open-source -- are typically static, pre-rendered, and lack generalizability \cite{shi2025towards,Brunetto_2023}, restricting the versatile auditory feedback necessary to model complex physical interactions. To overcome this data bottleneck, researchers have developed robust platforms such as ThreeDWorld \footnote{https://github.com/threedworld-mit/tdw} \cite{gan2021threedworldplatforminteractivemultimodal} and SoundSpaces 2.0 \footnote{https://soundspaces.org/} \cite{chen2023soundspaces20simulationplatform}. These state-of-the-art simulators generate multidirectional sound that is properly reverberated according to 3D scene geometry and (potentially) material composition. By utilizing experimentally-based Head Related Transfer Functions (HRTF), they simulate the average spectral cues of human hearing, enabling an agent to perceive accurately spatialized sound from sources within a 3D environment.
    
    These pipelines, however, are still primarily designed for other purposes, such as physical simulation and reinforcement learning, which introduces challenges when specifically generating scalable audio datasets. Thus, to address the practical difficulties of generating continuous, massive acoustic data for machine learning using these underlying platforms, we developed AudioWorldSim (available at \url{https://github.com/Luizerko/AudioWorldSim}), an open-source framework built on top of SoundSpaces 2.0. AudioWorldSim crucially resolves an underlying bug in the original SoundSpaces 2.0 continuous simulator that caused noticeable clicking sounds between steps -- unnatural artifacts that can be detrimental to machine learning models. Furthermore, by automating the rollout of random agent navigations, AudioWorldSim enables users to effortlessly generate audio datasets at scale across any compatible 3D scene. Ultimately, with world models in mind, the framework records the agent's actions at every simulation step. This preserves the vital action-consequence relationship of the agent interacting with the environment, capturing exactly how the sound changes in each ear accordingly -- essential for training world models to predict acoustic dynamics.

\section{The Simulator}

    Our simulator operates by placing a sound source and a virtual agent on the same floor level of a scene's navigation mesh (\texttt{NavMesh}). Users can define fundamental acoustic properties, such as the audio sampling rate, which defaults to 44.1 kHz to capture the full spectrum of human hearing. The environment supports both static, single-point Impulse Response (IR) computations and continuous temporal rollouts. In rollout mode -- the primary configuration for dataset generation -- the agent traverses the environment while the simulator continuously computes the evolving acoustic landscape based on a specified sequence of actions.

    Agent navigation is highly configurable to accommodate diverse modeling needs. In targeted navigation scenarios, the simulator autonomously samples valid, randomized starting and target points on the same floor, routing the agent toward the destination via the shortest path along the \texttt{NavMesh} -- a feature built on top of the tools from the Habitat-Sim \footnote{https://github.com/facebookresearch/habitat-sim} \cite{savva2019habitatplatformembodiedai}, a simulation environment used by SoundSpaces 2.0. Alternatively, manual navigation configurations provide explicit control over the agent's actions at each step.

    Because the underlying physics engine is disabled (following the methodology of SoundSpaces 2.0 \cite{chen2023soundspaces20simulationplatform}), agent movement operates kinematically as discrete spatial translations. Consequently, the temporal resolution of the simulation -- the time-step -- must be carefully calibrated. By default, a single forward action translates the agent by 0.2 meters, corresponding to a 0.2-second time-step. This parameter strictly governs the agent's perceived velocity. Variations in this hyperparameter directly impact how audio-based world models learn action-consequence dynamics, as it dictates the spatial delta between consecutive acoustic observations. Setting the time-step too high results in unnaturally slow movement, whereas setting it too low creates excessively fast traversal.

\begin{figure}[h]
        \centering
        \includegraphics[width=0.48\textwidth]{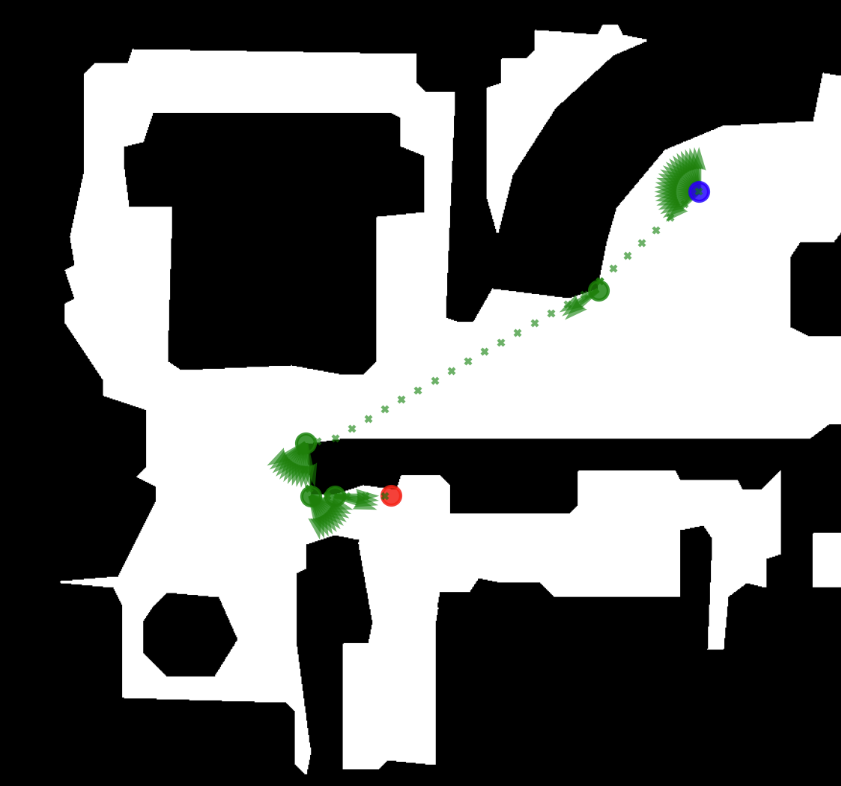}
        \hfill
        \includegraphics[width=0.43\textwidth]{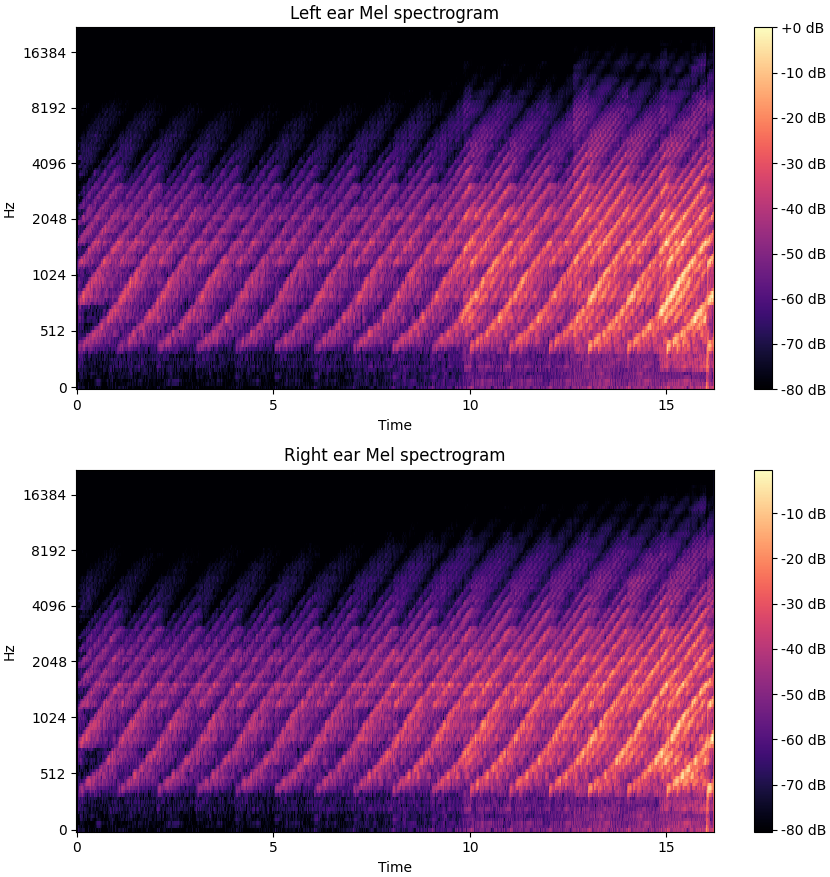}
        \caption{Visual and auditory breakdown of an agent's navigation. The physical trajectory along the \texttt{NavMesh} (left) illustrates the agent's path from its initial state (blue point) to the sound source (red point), guided by intermediate shortest-path waypoints (green points). Green crosses mark the actual discrete spatial steps taken along this route, while the arrows depict the sequence of rotational adjustments made to align the agent's orientation before it starts with translation again. This physical movement correlates directly with the acoustic energy captured in the left and right ear Mel spectrograms (right). Notice the interaural volume differences, such as the prominent vertical bands visible in the left ear spectrogram, when the agent rounds a corner.}
        \label{fig:nav_breakdown}
    \end{figure}

    For each simulation episode, the framework outputs a continuous binaural audio waveform, an exact sequence of discrete actions taken by the agent, and a comprehensive visual top-down map of the navigation trajectory. An optional video of the navigation can also be generated to facilitate the mapping of audio-visual outputs. Figure \ref{fig:nav_breakdown} illustrates this relationship, correlating the physical navigation path with the resulting spatialized auditory spectral features captured by the agent's virtual microphones. The rich acoustic interactions of our experiments, including the example depicted in the figure, were generated by leveraging the complex indoor geometries of 3D scenes from the Matterport3D \cite{chang2017matterport3dlearningrgbddata} and Replica \cite{replica19arxiv} datasets.

    Although not directly part of the simulator itself, our framework includes an automated audio processing pipeline triggered by default during large-scale dataset generation. This pipeline is dedicated to extracting ready-to-use spectral features, enabling users to directly apply machine learning algorithms -- particularly world models -- to either the entire audio sequence or discrete action steps. These processes are detailed in the following subsection.

    \subsection{Audio Processing Pipeline}

\begin{figure}[h]
            \centering
            \includegraphics[width=0.45\textwidth]{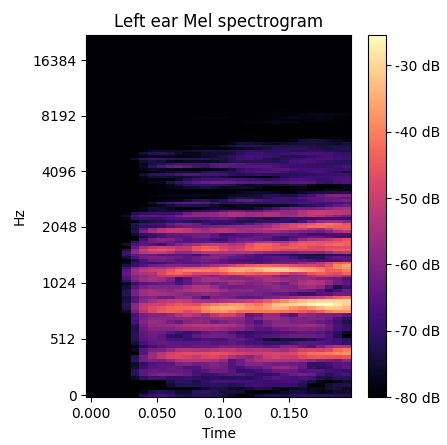}
            \hfill
            \includegraphics[width=0.45\textwidth]{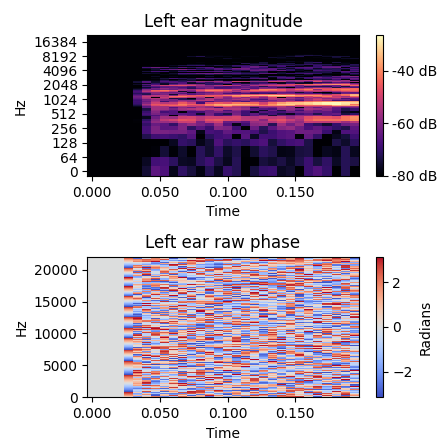}
            \caption{Comparison of output acoustic features segmented by discrete agent actions. A segment of the generated Mel spectrogram (left), which prioritizes spectral frequency resolution, and a segment of the raw STFT output (right), designed to preserve phase information essential for the spatial understanding capabilities of audio-based world models.}
            \label{fig:feature_extraction}
        \end{figure}

        The pipeline provides users with comprehensive feature sets: the complete Mel spectrogram of a run, the maximum power reference for the Mel spectrogram of a run, critical for audio reconstruction, and the complete raw Short Time Fourier Transform (STFT) output. Crucially for world models, it also outputs segmented versions of the spectrograms, sliced per agent action -- defaulting to a 0.2-second time-step. Figure \ref{fig:feature_extraction} visualizes these segmented features for both the Mel spectrogram and raw STFT formats. Note that any manual adjustments to the time-step or sample rate in the simulator must be identically updated in the audio processing pipeline to ensure accurate processing.

        To understand the design of our feature extraction, it is essential to consider how humans perceive spatial audio. Auditory localization relies on three primary mechanisms: Interaural Time Difference (ITD), which captures microsecond-level arrival delays between ears; Interaural Level Difference (ILD), which measures the disparity in acoustic intensity between ears caused by the head shadowing effect; and spectral cues provided by the Head-Related Transfer Function (HRTF), which filters frequencies based on torso and ear anatomy to disambiguate elevation and front/back locations.

        Because any standard sampling window spans a duration orders of magnitude larger than the microsecond-level temporal cues of ITD, preserving precise phase information is strictly necessary to capture these delays. However, standard Mel spectrograms inherently discard phase data, so we deliberately parameterize our Mel spectrograms to prioritize frequency resolution over temporal resolution. We compute these features using a window of 2048 samples and 128 frequency bands. Given the 44.1 kHz sample rate and a 0.2-second time-step, each action encompasses 8820 audio samples. Applying a hop length of 147 yields exactly 60 frames, resulting in a $128 \times 60$ Mel spectrogram per action. Note that if users modify the hop length parameter in this pipeline, they must ensure that the calculation $(\texttt{sample\_rate} \times \texttt{time\_step}) / \texttt{hop\_len}$ evaluates to an integer to maintain correct frame alignment.

        To explicitly capture ITD via phase information, ensuring posterior world models can have access to a more precise spatial understanding, the pipeline also generates raw STFTs. For this computation, we naturally reduce the window to 1024 samples to improve time resolution. Utilizing 512 frequency bands and maintaining the hop length of 147, this configuration generates a $512 \times 60$ raw spectrogram per action. Importantly, we preserve the raw complex numbers, natively encoding both amplitude and phase, and therefore allowing the the data to be utilized directly in complex-valued neural networks.

    \subsection{Scalable Dataset Generation}

        To support the massive data requirements of world models, AudioWorldSim incorporates a batch-processing pipeline designed for parallel execution at scale. During benchmark testing on a scene from the Matterport3D dataset, running 30 parallel workers on a 36-core Intel Core i9-10980XE processor consumed approximately 28 GB of RAM. This configuration generated around 1500 distinct trajectories -- ensured through strict seed control -- yielding roughly six hours of continuous, spatialized audio in under five hours of processing time.

        While the automated generation pipeline is highly efficient, it is subject to minor environmental edge cases. To maintain computational speed while resolving pathing issues, we implemented a depth-limited backtracking navigation algorithm. Despite this, approximately one percent of simulated trajectories experience navigation drift: cumulative imprecision in turning and forward movements causes the agent to become stuck against the collision mesh before reaching its target. However, these aborted runs still generate valid, albeit shorter, spatial audio sequences and corresponding action logs that remain perfectly viable for training. Furthermore, an estimated five percent of generation runs fail entirely due to inherent software faults in the underlying SoundSpaces 2.0 backend. Given the relatively rapid generation speed and the fact that these edge cases affect only a small fraction of total runs, massive, high-quality datasets can still be reliably compiled.

    \subsection{Resolving Framework Dependencies and Acoustic Artifacts}

        Our development addressed significant structural and dependency-related challenges within the original SoundSpaces 2.0 framework. The official installation guide relies on a headless version of Habitat-Sim, whereas executing their continuous simulator demands the full software suite. Furthermore, the pipeline depends on specific, deprecated versions of both Habitat-Sim and Habitat-Lab \footnote{https://github.com/facebookresearch/habitat-lab} \cite{savva2019habitatplatformembodiedai}, making environment resolution notoriously difficult. To bypass this bottleneck, we isolated the core sound simulation backend -- which natively handles audio spatialization for isolated agent-source pose pairs -- and built our minimal, targeted simulator directly on top of it. This allowed us to engineer a robust continuous navigation framework that queries the acoustic dynamics independently, entirely removing the reliance on outdated agent-simulation dependencies.
        
        Beyond structural decoupling, we implemented a critical correction to how SoundSpaces 2.0 synthesizes continuous audio across discrete simulation steps. To generate continuous audio, the system must transition smoothly between discrete spatial locations. At a given simulation step from time $t$ to $t+1$, the framework retrieves the acoustic IR for both the previous pose at $t$ and the new pose at $t+1$. We then convolve both of these IRs with the source audio segment corresponding to time $t+1$. The first convolution represents the acoustic state if the agent had remained stationary, while the second represents the state at the new location. A crossfade is then applied between the two to create the transition. Conceptually, this blends the initial audio at the new pose with the expected acoustic profile of the previous pose, approximating the perceptual experience of continuous spatial movement. The original SoundSpaces implementation erroneously convolved the IR at pose $t$ with the source audio segment from time $t$ rather than $t+1$, effectively mixing present and past audio segments. This introduced a temporal discontinuity that our framework explicitly corrects.

        More importantly, we resolved a severe audio artifact: a noticeable clicking sound that plagued the original continuous simulator. The root cause of this artifact lies in the varying lengths of IRs across different spatial positions, which fluctuate based on local acoustic reverberation. Without proper zero-padding to equalize the lengths of shorter IRs against longer ones, the resulting convolved audio segments become temporally misaligned during the crossfade. Specifically, if the past IR is longer than the current IR, the convolution effectively delays the past audio, shifting it further into the future. When the crossfade concludes and the system snaps entirely to the current IR, this temporal misalignment causes a sudden audio shift that manifests as a harsh click.

\begin{figure}[h]
            \centering
            \hfill
            \includegraphics[width=0.45\textwidth]{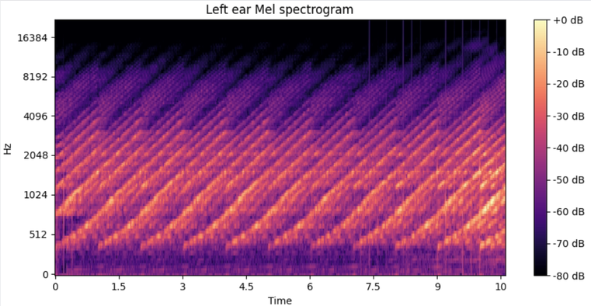}
            \hfill
            \includegraphics[width=0.45\textwidth]{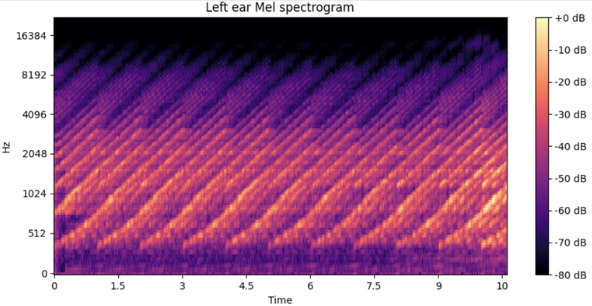}
            \hfill
            \caption{Mel spectrogram comparison of continuous audio synthesis. In the original implementation (left), temporal misalignment between unequal IRs causes vertical artifact lines (clicking sounds) between some simulation steps. Our corrected AudioWorldSim implementation utilizes proper zero-padding (right), resulting in a artifact-free audio continuum.}
            \label{fig:crossfade_fix}
        \end{figure}

        This phenomenon can be heard throughout most of the original SoundSpaces 2.0 demonstration video \footnote{https://www.youtube.com/watch?v=4uiptTUyq30\&feature=youtu.be}, becoming particularly pronounced as the agent nears the sound source. As the distance to the source decreases, the proportion of direct sound increases while reverberation decreases. Consequently, the new IRs become significantly shorter than the preceding ones, exacerbating the temporal misalignment. This not only amplifies the harshness of the clicking artifact, but also increases its frequency, as every successive step toward the source generates a newly misaligned, shorter IR. By implementing proper zero-padding, our framework effectively aligns the segments and eliminates the issue. The efficacy of this correction is visually evident in Figure \ref{fig:crossfade_fix}, which compares the resulting Mel spectrograms before and after our fix.

\section{Conclusion}

High-quality audio datasets, particularly those that are open-source, binaural, and feature realistic acoustic propagation tailored for spatial navigation rather than speech, are exceedingly rare. The few existing datasets are typically pre-rendered and lack generalizability \cite{shi2025towards,Brunetto_2023}, restricting the versatile and scalable auditory feedback necessary to model more complex physical interactions. Furthermore, while realistic acoustic simulators such as ThreeDWorld \cite{gan2021threedworldplatforminteractivemultimodal} and SoundSpaces 2.0 \cite{chen2023soundspaces20simulationplatform} are available, they are primarily optimized for active reinforcement learning and physical simulation rather than streamlined dataset extraction. To address this gap, we introduce AudioWorldSim, an open-source, highly scalable, and easily configurable framework designed to leverage these simulation capabilities to generate continuous acoustic data for spatial navigation and broader world understanding.

By resolving critical dependencies and temporal audio bugs in the underlying SoundSpaces 2.0 \cite{chen2023soundspaces20simulationplatform}, we achieved a robust pipeline capable of generating artifact-free spatial audio. This work democratizes access to high-fidelity audio data, giving researchers the power to generate massive datasets tailored to their specific needs, utilizing any compatible 3D scene. Created with modern machine learning paradigms in mind -- particularly audio-based world models -- AudioWorldSim preserves the vital action-consequence relationship of an agent interacting with its environment, ultimately aiming to accelerate and expand the field of audio-based artificial intelligence.

\section{Limitations and Future Work}

While AudioWorldSim significantly streamlines continuous audio dataset generation, several limitations remain. First, the version of SoundSpaces 2.0 \cite{chen2023soundspaces20simulationplatform} that our framework builds upon currently exhibits unresolved issues regarding material-based acoustic computations. Enabling material-specific sound propagation can compromise the acoustic quality and consistency across multiple simulation seeds. As we have not yet resolved this upstream issue, we advise against using the material propagation feature for large-scale dataset generation. Second, although we successfully decoupled the continuous simulator from outdated agent-simulation dependencies, preparing new 3D environments still requires adhering to the spatial and metadata configurations mandated by the SoundSpaces 2.0 backend.

Furthermore, our initial goal was to publicly release a massive, pre-rendered acoustic dataset to allow researchers to experiment immediately without running the simulation pipeline themselves. However, strict licensing agreements associated with the Matterport3D \cite{chang2017matterport3dlearningrgbddata} and Replica \cite{replica19arxiv} datasets prevented us from distributing the generated audio. As a mitigation, we patched the SoundSpaces 2.0 script to automatically download, configure, and utilize a sample scene from Matterport3D. This ensures that users can generate a comprehensive dataset for at least one complex environment directly out of the box. Consequently, a highly valuable direction for future work would be for an entity with the appropriate dataset licenses to leverage AudioWorldSim to generate and publicly distribute a large-scale, multi-scene open dataset for the broader machine learning community.

Looking ahead, we also identify two major technical avenues for future work. Developing an interactive version of the simulator is a natural next step, which would significantly facilitate rapid testing and debugging of acoustic configurations. Additionally, deeper integration with the native RGB-D visual sensors of SoundSpaces 2.0 would allow the simultaneous extraction of synchronized visual data, enabling the creation of multimodal datasets for audio-visual world models.

\section{Ethical Statement and AI Usage Declaration}

There are no specific ethical concerns, human subject privacy risks, or safety considerations associated with this work. All experiments and datasets are generated entirely within synthetic simulation environments.

Regarding artificial intelligence assistance, Gemini 3.1 chat \cite{GoogleDeepMind2026Gemini31} was utilized during the development of this project to assist with code generation, with direct inclusion of various code snippets created by the AI upon request. Gemini 3.1 chat was also employed during the writing of this technical report. All core conceptual ideas, technical explanations, and an entire draft were human-written, then the AI was utilized for text restructuring and refinement, as well as language revision. Finally, a thorough human revision of every single paragraph took place to get to this final version.

\bibliographystyle{alpha}
\bibliography{sample}

\end{document}